\documentclass[12pt]{article}
\pdfoutput=1    
  
\usepackage{hyperref}    
\usepackage{amsmath}      
\usepackage{amssymb} 
\usepackage{graphicx}
\usepackage{url}
\usepackage{enumerate}
 
\usepackage{float,wrapfig,color} 

\allowdisplaybreaks
\def\eq#1{(\ref{#1})}
\def\s[#1\s]{\begin{align}\begin{split}#1\end{split}\end{align}}
\def\[#1\]{\begin{align}#1\end{align}}

\def\bpsi{{\bar\psi}}
\def\bvphi{\bar \varphi}
\def\vphi{{\varphi}}
\def\pvphi{\varphi_\parallel}
\def\pbpsi{\bpsi_\parallel}
\def\pvphi{\vphi_\parallel}
\def\ppsi{\psi_\parallel}
\def\pbvphi{\bvphi_\parallel}
\def\tbpsi{\bar \psi_\perp}
\def\tpsi{\psi_\perp}
\def\tvphi{\varphi_\perp}
\def\tbvphi{\bar \varphi_\perp}

\begin{document}

\begin{titlepage} 

\title{
 \hfill\parbox{4cm}{ \normalsize YITP-22-98}\\   
\vspace{1cm} 
Real tensor eigenvalue/vector distributions \\ of the Gaussian tensor model 
via a four-fermi theory}

\author{Naoki Sasakura\footnote{sasakura@yukawa.kyoto-u.ac.jp}
\\
{\small{\it Yukawa Institute for Theoretical Physics, Kyoto University, }}\\
{\small {\it and } } \\
{\small{\it CGPQI, Yukawa Institute for Theoretical Physics, Kyoto University,}} \\
{\small{\it Kitashirakawa, Sakyo-ku, Kyoto 606-8502, Japan}}
}


\maketitle 

\begin{abstract}  
Eigenvalue distributions are important dynamical quantities in matrix models, and it is an interesting challenge to 
study corresponding quantities in tensor models.  
We study real tensor eigenvalue/vector distributions for real symmetric order-three random tensors with the Gaussian distribution
as the simplest case.
We first rewrite this problem as the computation of a partition function of a four-fermi theory
with $R$ replicated fermions. 
The partition function is exactly computed for some small-$N,R$ cases, 
and is shown to precisely agree with Monte Carlo simulations.
For large-$N$, it seems difficult to compute it exactly, 
and we apply an approximation using a self-consistency equation for two-point functions 
and obtain an analytic expression.
It turns out that the real tensor eigenvalue distribution obtained by taking $R=1/2$ 
is simply the Gaussian within this approximation.
We compare the approximate expression with Monte Carlo simulations, 
and find that, if an extra overall factor depending on $N$ is multiplied 
to the the expression, it agrees well with the Monte Carlo results.
It is left for future study to improve the approximation for large-$N$
to correctly derive the overall factor.
\end{abstract}
\end{titlepage}
 
\section{Introduction}
\label{sec:intro}
Eigenvalue distributions are important dynamical quantities in matrix models. 
The distributions give quantitative/qualitative insights into complex dynamical systems 
with strongly coupled degrees of freedom, such as atomic systems \cite{Wigner}. 
Topological transitions of the distributions  \cite{Gross:1980he,Wadia:1980cp} 
give important insights into the properties of gauge theories and 
two-dimensional quantum gravity, such as phase structures in particular. 
The distributions are also important as one of the main tools to solve the matrix models \cite{Brezin:1977sv}.  
  
Considering recent attentions to tensor models \cite{Ambjorn:1990ge,Sasakura:1990fs,Godfrey:1990dt,Gurau:2009tw} 
it would be interesting to study corresponding quantities and their roles in tensor models. 
While, in various 
applications \cite{qibook}\footnote{See for example \cite{Ouerfelli:2022rus} for the tensor model perspectives on such applications.}, 
it is important to develop techniques to obtain eigenvalues/vectors \cite{Qi,lim,cart} for given tensors,
it is rather more important to understand statistical properties of tensor eigenvalues/vectors
for ensembles of tensors in tensor models, where tensors are dynamical.
In fact, not many results are known about their statistical properties: 
The expectation values of the numbers of real eigenvalues are computed for random real tensors \cite{realnum1,realnum2};
the sizes of the largest eigenvalues are estimated for random tensors \cite{Evnin:2020ddw};
Wigner semi-circle law in matrix models has been extended to tensor models \cite{Gurau:2020ehg}. 

A basic question about eigenvalues/vectors in tensor models is their distributions. 
In a previous study  \cite{Sasakura:2022zwc}, the present author derived an exact 
formula\footnote{The formula is expressed with the confluent hypergeometric function of the second kind.}
for signed distributions of real eigenvectors for real symmetric order-three random 
tensors with the Gaussian distribution: 
Each eigenvector contributes to the distribution by $\pm1$ depending on the 
sign of an associated Hessian matrix.
An interesting intermediate step 
was that the problem was rewritten as the computation of a partition function of a four-fermi theory. 
This was achieved by Gaussian integrations over bosonic variables after rewriting the problem 
as a partition function of a system of bosonic and fermionic variables.
The final formula was obtained by exactly computing the partition function of the four-fermi theory. 

The above procedure of the previous paper can be extended to other kinds of eigenvalue/vector distributions. 
The purpose of the present paper is to extend it to the distribution of real eigenvalues/vectors for real symmetric 
order-three random tensors with the Gaussian distribution. In the previous paper, we only dealt with one couple of fermions, but, 
in this paper, we deal with $R$ replicas of fermions, and the eigenvector/value distribution is supposed to 
be obtained by an analytic continuation to $R=1/2$.
We derive a four-fermi theory with the replicated fermions, which is more complex than that of the previous paper.
We exactly compute the partition function of the four-fermi theory for some small-$N,R$ cases and show the
precise agreement with Monte Carlo results.
On the other hand, for large-$N$, we apply an approximation using the Schwinger-Dyson equation (See \ref{app:sd}), and obtain the eigenvalue/vector distribution by formally taking $R = 1/2$.
We find that the obtained approximate analytic expression has good agreement with Monte Carlo results, 
if we multiply an extra overall factor depending on $N$ to the expression, 
where $N$ denotes the range of the tensor indices.
 
This paper is organized as follows. 
In Section~\ref{sec:dist}, we define the distribution of the real tensor eigenvalues/vectors of
the Gaussian tensor model, and rewrite the distribution as a partition function of a system of bosonic and fermionic variables.
In Section~\ref{sec:fourfermi}, by Gaussian integrations over the bosonic variables of the system, we obtain 
a four-fermi theory. 
In Section~\ref{sec:exact}, we exactly compute the partition function of the four-fermi theory for some small-$N,R$ cases.
In Section~\ref{sec:approximation}, we compute the partition function of the four-fermi theory 
for large-$N$ by applying an approximation using the
Schwinger-Dyson equation for two-point functions, and obtain an approximate 
analytic expression of the eigenvalue/vector distribution by formally taking $R=1/2$.
In Section~\ref{sec:monte}, we perform some Monte Carlo simulations. The exact computations for small-$N,R$
are shown to precisely agree with the Monte Carlo results.
On the other hand,
we find that the approximate analytic expression of the eigenvalue/vector distribution 
agrees well with the Monte Carlo result, if it is multiplied by an extra factor depending on $N$.
The final section is devoted to a summary and future problems.  
In \ref{app:relation}, we explain an overlap between the main result of Section~\ref{sec:approximation}
and a known result from random matrix theory.

\section{Real tensor eigenvalue/vector distributions}
\label{sec:dist}
In this paper, we restrict ourselves to the symmetric real tensors with three indices, 
$C_{abc}=C_{bac}=C_{bca}\in \mathbb{R} \ (a,b,c=1,2,\ldots,N)$, as the simplest case.
There are several different definitions of tensor eigenvalues/vectors \cite{Qi,lim,cart}. In this paper we employ
the definition that real eigenvectors of a given $C$ are defined by $v$'s satisfying
 \[
C_{abc} v_b v_c = v_a,\ v\in \mathbb{R}^N.
\label{eq:eg}
\]
Here repeated indices are assumed to be summed over, as will also be assumed throughout this paper.
Then the real eigenvector distribution for a given $C$ is given by
\[
\rho(v,C)=\sum_{i=1}^{N_{\rm sol}(C)} \prod_{a=1}^N \delta(v_a-v_a^i),
\label{eq:defrhoC}
\]
where $v^i\ (i=1,2,\ldots,N_{\rm sol}(C))$ are all the solutions of \eq{eq:eg}.

The expression \eq{eq:defrhoC} can be rewritten as 
\[
\rho(v,C)=\left | \det M \right |\, \prod_{a=1}^N \delta\left( v_a -C_{abc} v_b v_c \right),
\label{eq:defrho}
\]
where $\det$ represents taking the determinant, $|\cdot |$ is to take the absolute value,  and $M$ is a matrix defined by
\[
M_{ab}=\frac{\partial}{\partial v_a} \left( v_b -C_{bcd} v_c v_d\right) =\delta_{ab} -2 C_{abc} v_c.
\label{eq:defM}
\]

Suppose the tensor $C$ is randomly distributed with the Gaussian distribution.
Then the distribution of $v$ is given by
\[
\rho(v) = \langle \rho(v,C) \rangle_C = A^{-1} \int_{\mathbb{R}^{D_C}} dC\, e^{-\alpha C^2}\, \rho(v,C),
\label{eq:rhov}
\]
where $dC=\prod_{a\leq b \leq c=1}^N dC_{abc}$,
$C^2=C_{abc} C_{abc}$, $\alpha$ is a positive real number,  $A= \int_{\mathbb{R}^{D_C}} dC\, e^{-\alpha C^2}$, 
and $D_C=N(N+1)(N+2)/6$, i.e., the total number of independent components of $C$.

In the previous paper \cite{Sasakura:2022zwc}, we computed a similar quantity which has $\det M$ instead of $|\det M|$ in \eq{eq:defrho}. 
The previous quantity was easier to compute, since it can simply be expressed by a couple of 
fermions by using the formula, $\det M=\int d\bpsi d\psi \, e^{\bpsi_a M_{ab} \psi_b}$ \cite{zinn}.
On the other hand,
what makes the expression \eq{eq:defrho} difficult to deal with
is that $|\det M|$ is not analytic. One way to turn it to an analytic expression is to consider 
a quantity, 
\[ 
\rho(v,C,R,\epsilon) = \left( \det \left(M^2+\epsilon I \right)\right)^R  \prod_{a=1}^N \delta\left( v_a -C_{abc} v_b v_c \right),
\]
where $\epsilon>0$ is a regularization parameter, and $I$ is the identity matrix, $I_{ab}=\delta_{ab}$. 
Note that the regularization parameter keeps the $O(N)$ invariance of the 
system\footnote{The system is invariant under the orthogonal transformation in the vector space 
associated to the lower index.}. 
Then, for random $C$ with the Gaussian distribution, we have 
\[
\rho(v,R,\epsilon)=\langle \rho(v,C,R,\epsilon)\rangle _C=A^{-1}\int_{\mathbb{R}^{D_C}} dC\, e^{-\alpha C^2}
\left( \det \left(M^2+\epsilon I \right)\right)^R  \prod_{a=1}^N \delta\left( v_a -C_{abc} v_b v_c \right).
\label{eq:rhoreg}
\]
The expression corresponding to \eq{eq:rhov} can be obtained by putting $R=1/2$ and
taking the $\epsilon \rightarrow +0$ limit:
\[
\rho(v)=\rho(v,1/2,+0).
\label{eq:rhovhalf}
\]
As we will find in Section~\ref{sec:approximation}, the regularization parameter $\epsilon$ is necessary for 
the large-$N$ approximate computation,
to unambiguously obtain an expression 
which is valid in all the region of $v$:
Directly starting from the expression with $\epsilon=0$ has a singularity 
and it is not clear how the expression should be extended to the whole region.

The real eigenvalues accompanied with real eigenvectors
(Z-eigenvalues in the terminology of \cite{Qi}) for a real symmetric order-three tensor $C$ are the $z$'s satisfying
\[
C_{abc}w_b w_c= z\, w_a,
\label{eq:zeigeq}
\]
where $z \in \mathbb{R}$, $w\in \mathbb{R}^N$with $|w|=1$ ($|w|=\sqrt{w_a w_a}$). 
The relation to \eq{eq:eg} is $w=v/|v|$ and $z=1/|v|$.
Therefore, once we obtain $\rho(v)$, the Z-eigenvalue distribution is obtained by 
\s[
\rho_{\rm eigenvalue}(z)&=\rho(v)S^{N-1} |v|^{N-1} \frac{d|v|}{dz} \\
&=\rho(1/z)S_{N-1} z^{-N-1} , 
\label{eq:rhoeigen} 
\s]
where $S_{N-1}=2 \pi^{N/2}/\Gamma(N/2)$ is the surface volume of a unit sphere in an $N$-dimensional space.
To derive this expression, we have used that $\rho(v)$ actually depends only on $|v|$, because of the $O(N)$-invariance of \eq{eq:rhov}, and have used $d^Nv=S_{N-1} |v|^{N-1} d|v|$.
Note that we have abusively written $1/z$ as the argument on the righthand side of \eq{eq:rhoeigen} to
represent an arbitrary vector of size $1/z$.

\section{A four-fermi theory with $R$ replicas}
\label{sec:fourfermi} 
By using the well-known formulas, $\int_\mathbb{R} dx\, e^{i xy}=2 \pi \delta(y)$ and $\int d\bpsi d\psi \, e^{\bpsi_a T_{ab} \psi_b}=\det T$ 
\cite{zinn}, the expression \eq{eq:rhoreg} can be rewritten as
\[
\rho(v,R,\epsilon)=A^{-1} (2 \pi)^{-N} \int dC d \lambda d\bar\psi d\psi \, e^{S_1},
\label{eq:zfb}
\]
where 
\[
S_1=-\alpha C^2 + i \lambda_a (v_a -C_{abc}v_b v_c) +\bpsi_a^i M^2_{ab} \psi_{bi}+
\epsilon \bpsi_a^i \psi_{ai}.
\label{eq:sfb}
\]
Here we have assumed $R$ to be integer, $\bar \psi_a^i,\psi_{ai}\ (a=1,2,\ldots,N;\ i=1,2,\ldots,R)$ are fermionic, 
and $\lambda_a \ (a=1,2,\ldots,N)$ are bosonic with the integration region $\mathbb{R}^N$.  
Note that the paired new 
indices (namely, $i$) are also assumed to be summed over in the above expression, as will be assumed throughout this paper.
Note also that the system \eq{eq:zfb} is invariant under the following 
$GL(R)$ transformation concerning the new index:
\s[
\psi'{}_{ai}&=G{}_i{}^{i'} \psi_{ai'},\  \bar \psi'{}_a^i= \bar \psi^{i'}_{a} G^{-1}{}_{i'}{}^{i},
\s]
for $G \in GL(R)$. 
Another comment is 
that the appearance of the imaginary number in \eq{eq:sfb} does not violate the reality of the integral, since 
the integral is symmetric under $\lambda\rightarrow -\lambda$.

In the similar manner as the procedure taken in the previous paper \cite{Sasakura:2022zwc}, 
we want to first integrate over the bosonic variables, $C$ and $\lambda$, to obtain
a fermionic theory.  However, $M^2$ contains $C$ in a quadratic manner and is not straightforward to deal with. 
To circumvent this matter, let us introduce another pair of fermions $\bar \varphi,\varphi$ to rewrite \eq{eq:sfb} 
into an expression linear in $M$:
\[
\rho(v,R,\epsilon)=A^{-1} (2 \pi)^{-N}(-1)^{NR} \int dC d \lambda d\bar\psi d\psi d\bar\varphi d\varphi \, e^{S_2},
\label{eq:newrho}
\]
where 
\s[
S_2=-\alpha C^2 + &i \lambda_a (v_a -C_{abc}v_b v_c) -\bar\varphi_a^i \varphi_{ai} + \epsilon \bar \psi_a^i \psi_{ai}
 -\bar \psi_a^i M_{ab} \varphi_{bi}-\bar\varphi_a^i M_{ab} \psi_{bi}.
 \label{eq:defofs}
\s]
The equivalence between \eq{eq:zfb} and  \eq{eq:newrho} can be shown by
\s[
\int d\bar \varphi d\varphi\,  e^{-\bar\varphi_a \varphi_a  -\bar \psi_a M_{ab} \varphi_b-\bar\varphi_a M_{ab} \psi_b}&=
\int d\bar \varphi d\varphi\,  e^{-(\bar\varphi_a +\bar \psi_b M_{ba})(\varphi_a+ M_{ac} \psi_c)+\bar \psi_a M^2_{ab} \psi_b} \\
&=(-1)^N e^{\bar \psi_a M^2_{ab} \psi_b}.
\s]

Now let us first integrate over $C$, assuming that the final result does not depend on this change of the order of the integrations. 
The terms containing $C$ in \eq{eq:defofs} are
\[
S_C=-\alpha C^2 -i  C_{abc}\lambda_a v_b v_c +2 C_{abc} \bar\psi_a^i \varphi_{bi} v_c 
+2 C_{abc} \bar\varphi_a^i \psi_{bi} v_c.
\]
Therefore the integration over $C$ is just a Gaussian integration and we obtain
\[
\int_{\mathbb{R}^{D_C}} dC e^{S_C}=A\, e^{\delta S_C},
\label{eq:cint}
\]
where
\s[
\delta S_C&=\frac{1}{\alpha} \left( \frac{1}{6}\sum_{\sigma} \left(
-\frac{i}{2}  \lambda_{\sigma_a} v_{\sigma_b} v_{\sigma_c}  +  \bar\psi^i_{\sigma_a} \varphi_{\sigma_b i} v_{\sigma_c} 
+ \bar\varphi^i_{\sigma_a} \psi_{\sigma_b i} v_{\sigma_c}\right)
\right)^2.
\label{eq:sc1}
\s]
Here the summation over $\sigma$ is over all the permutations of $a,b,c$, 
the necessity of which comes from that $C$ is a symmetric tensor. 
Explicitly expanding the expression \eq{eq:sc1}, we obtain
\[
\delta S_C =-\frac{|v|^4}{12 \alpha} \lambda_a B_{ab} \lambda_b - i \lambda_a D_a+E
\label{eq:sc}
\]
with
\[
&B_{ab}=\delta_{ab} +2 \hat v_a \hat v_b=I_{\perp\,ab}+3 I_{\parallel\,ab}, \\
&D_a=\frac{|v|^3}{3 \alpha} \left( \pbpsi^i \pvphi{}_i  +\pbvphi^i \ppsi{}_i \right)\hat v_a +
\frac{|v|^3}{3 \alpha} \left( 
\bpsi_a^i \pvphi{}_i+\pbpsi^i \vphi_{ai}+\bvphi_a^i \ppsi{}_i+\pbvphi^i \psi_{ai}
\right),
\label{eq:defofd}  \\
&E=\frac{1}{\alpha}\left( 
\frac{1}{6} \sum_\sigma \left( 
\bpsi_{\sigma_a}^i \vphi_{\sigma_bi} v_{\sigma_c}
+\bvphi_{\sigma_a}^i \psi_{\sigma_bi} v_{\sigma_c} 
\right)
\right)^2,
\]
where $\hat v_a=v_a/|v|$, and $\parallel$ denotes the projection to $\hat v$, namely, $\psi_\parallel=\psi_a \hat v_a$, and so on.
As above, the matrix $B$ can be rewritten as a linear sum of the projection operators, where $I_\parallel$ is the projection operator to  
the one-dimensional linear subspace spanned by $\hat v$, 
and $I_\perp$ is to the subspace transverse to $\hat v$.

From \eq{eq:defofs} and \eq{eq:sc}, the remaining terms containing $\lambda$ are
\[
S_\lambda=-\frac{|v|^4}{12 \alpha} \lambda_a B_{ab} \lambda_b + i \lambda_a (v_a-D_a).
\]
Integrating over $\lambda$ gives
\[
\int_{\mathbb{R}^N} d\lambda\ e^{S_\lambda} =(12 \pi \alpha)^\frac{N}{2} |v|^{-2 N} (\det B)^{-\frac{1}{2}} e^{\delta S_\lambda}
 \]
 with
\s[
\delta S_\lambda&=- 3 \alpha |v|^{-4} (v_a-D_a) B^{-1}_{ab} (v_b-D_b)\\
& =-3 \alpha |v|^{-4} \left( D_\perp \cdot D_\perp + \frac{1}{3} (|v|-D_\parallel)^2 \right)\\
&=-\frac{\alpha}{|v|^2} + 2 \left(\pbpsi^i \pvphi^i+ \pbvphi^i \ppsi^i \right) -3 \alpha |v|^{-4}\left( D_\perp \cdot D_\perp +
\frac{1}{3} D_\parallel^2 
\right),
\label{eq:deltaslam}
\s]
where we have used $B^{-1}=I_\perp + \frac{1}{3} I_\parallel$, and from \eq{eq:defofd} 
the projections of $D$ to the parallel and transverse
directions to $\hat v$ are respectively given by
\s[
D_\perp&=\frac{|v|^3}{3 \alpha} \left( \tbpsi^i \pvphi{}_i+\pbpsi^i \tvphi{}_i+\tbvphi^i \ppsi{}_i+\pbvphi^i \tpsi{}_i\right), \\
D_\parallel &= \frac{|v|^3}{\alpha} \left(\pbpsi^i \pvphi{}_i+ \pbvphi^i \ppsi{}_i \right).
\s]
While the last term of \eq{eq:deltaslam} gives some four-fermi interaction terms, the second term
gives some corrections to the quadratic terms in the parallel direction. 

Collecting all the results above and using $\det B=3$, we obtain a four-fermi theory,
\[
\rho(v,R,\epsilon)=3^\frac{N-1}{2} \pi^{-\frac{N}{2}} \alpha^\frac{N}{2} |v|^{-2 N} e^{-\frac{\alpha}{|v|^2}}
(-1)^{NR} \int d\bpsi d\psi d\bvphi d\vphi \, e^{K+K_\parallel+V},
\label{eq:fourfermi3}
\]
where the quadratic term $K,\, K_\parallel$ and the four-fermi interaction terms $V$ are respectively given by
\s[
K&=
\left(
\begin{array}{cc}
\tbpsi^i & \tbvphi ^i 
\end{array}
\right)
\left(
\begin{array}{cc}
\epsilon & -1 \cr
-1& -1  
\end{array}
\right)
\left(
\begin{array}{c}
\tpsi{}_i \cr \tvphi{}_i  
\end{array}
\right),  \\
K_{\parallel}&=
\left(
\begin{array}{cc}
\pbpsi^i & \pbvphi^i 
\end{array}
\right)
\left(
\begin{array}{cc}
\epsilon & 1 \cr
1& -1  
\end{array}
\right)
\left(
\begin{array}{c}
\ppsi{}_i \cr \pvphi{}_i  
\end{array}
\right),  \\
V&=-3 \alpha |v|^{-4}\left( D_\perp \cdot D_\perp +\frac{1}{3} D_\parallel^2
\right)+E   \\
&=-\frac{|v|^2}{6 \alpha} \Big(
\tbpsi^i\cdot \tbpsi^j \tvphi{}_i\cdot \tvphi{}_j+\tbpsi^i\cdot \tvphi{}_j \tbpsi^j\cdot \tvphi{}_i
+\tbvphi^i\cdot \tbvphi^j \tpsi{}_i\cdot \tpsi{}_j \\
&\ \ \ \ \ \ \ \ \ \ \ \ 
+\tbvphi^i\cdot \tpsi{}_j \tbvphi^j\cdot \tpsi{}_i  +2 \tbpsi^i\cdot \tbvphi^j \tvphi{}_i\cdot \tpsi{}_j+2 \tbpsi^i\cdot \tpsi{}_j \tbvphi^j\cdot \tvphi{}_i \Big).
 \label{eq:vint}
\s]  

What is surprising in $V$ in \eq{eq:vint} is that the parallel components are not contained. Therefore,
the parallel components are free theory and can trivially be integrated over:
\[
\int d\pbpsi d\ppsi d\pbvphi d\pvphi \, e^{K_{\parallel}}=(-1)^R,
\]
where we have already taken the $\epsilon\rightarrow +0$ limit, since this is smooth.
Then we finally obtain a four-fermi theory containing only the transverse components,
\[
\rho(v,R,\epsilon)=3^\frac{N-1}{2} \pi^{-\frac{N}{2}} \alpha^\frac{N}{2} |v|^{-2 N} e^{-\frac{\alpha}{|v|^2}}
(-1)^{(N-1)R} \int d\tbpsi d\tpsi d\tbvphi d\tvphi \, e^{K+V}.
\label{eq:fourfermi}
\]

\section{Exact computations for small-$N,R$}
\label{sec:exact}
Simple crosschecks of the formula, \eq{eq:fourfermi} with \eq{eq:vint}, 
can be performed by exactly computing it.
In the $N=1$ case, we have no transverse directions, and hence
\eq{eq:fourfermi} gives
\[
\rho_{N=1} (v,R,0)=\pi^{-\frac{1}{2}} \alpha^\frac{1}{2} v^{-2} e^{-\frac{\alpha}{|v|^2}}.
\label{eq:rhodv}
\] 
Indeed, from that the eigenvector is given by $v=1/C$ for $N=1$, the Gaussian distribution 
of $C$ is the same as \eq{eq:rhodv}, because
$\pi^{-\frac{1}{2}} \alpha^\frac{1}{2} e^{-\alpha C^2} dC=\rho_{N=1} (v,R,0) dv$.

Another obvious consequence is that, for integer $R$, the four-fermi theory \eq{eq:fourfermi} 
can in principle be computed by expanding the exponential in $K, V$:
\[
\int d\tbpsi d\tpsi d\tbvphi d\tvphi \, e^{K+V}=\sum_{n=0}^{(N-1)R} \frac{1}{n! (2 (N-1)R-2n
)!} \int d\tbpsi d\tpsi d\tbvphi d\tvphi \, V^n\, K^{2 (N-1)R-2 n}.
\label{eq:expansion}
\]  
This formula comes from the fact that each term of the integrand must be a product of exactly $4(N-1)R$ fermions 
for the fermionic integral to be non-zero \cite{zinn}.
The $\epsilon \rightarrow +0$ limit of \eq{eq:expansion} is straightforward, and,  for integer $R$,
\eq{eq:fourfermi} generally has the form,
\[
\rho(v,R,0)=3^\frac{N-1}{2} \pi^{-\frac{N}{2}} \alpha^\frac{N}{2} |v|^{-2 N} e^{-\frac{\alpha}{|v|^2}} {\cal L}_{N,R},
\label{eq:rhol}
\] 
where ${\cal L}_{N,R}$ is a polynomial function of $x=|v|^2/(6\alpha)$, ${\cal L}_{N,R}=1+c_1 x +c_2 x^2+\cdots+c_{(N-1)R} x^{(N-1)R}$,
where $c_i$ are some coefficients depending on $N,R$. 
Then, as in \eq{eq:rhovhalf}, $\rho(v)$ could 
be obtained by taking the analytic continuation of ${\cal L}_{N,R}$ to $R=1/2$.

However, the exact computation of ${\cal L}_{N,R}$ for general $N,R$ seems to be a difficult task. 
It becomes more and more cumbersome very quickly, as $N,R$ increase.    
Still it is doable for some small $N,R$ by using computers.
By a Mathematica package for Grassmann variables \cite{grassmann}, we obtain
\s[
{\cal L}_{N=2,R=1}&=1+4 x, \\
{\cal L}_{N=3,R=1}&=1+4x+28x^2, \\
{\cal L}_{N=2,R=2}&=1+24 x+48 x^2, \\
{\cal L}_{N=3,R=2}&=1 + 40 x + 552 x^2 + 1248 x^3 + 5136 x^4.
\label{eq:nreq}
\s]
The formula \eq{eq:rhol} with \eq{eq:nreq} will be checked by Monte Carlo simulations in Section~\ref{sec:monte}.

\section{Approximate computations for large-$N$}
\label{sec:approximation}

As was demonstrated in the last part of Section~\ref{sec:exact}, the expression \eq{eq:fourfermi} 
can in principle be computed for general integer
$N,R$, and $R=1/2$ could be taken by analytic continuation. However, this does not seem
straightforward in practice. 
 Therefore, in this section, we apply the Schwinger-Dyson equation to the four-fermi theory, assuming the 
factorization of the four-point correlation functions into the products of two-point functions. 
A self-contained brief explanation of the approximation is given in \ref{app:sd}.

The four-fermi theory \eq{eq:fourfermi} has the following symmetry, $O(N-1)\times GL(R)$,
for the transverse variables $X,\, \bar X \ (X=\psi_\perp,\varphi_\perp)$:
\s[
X'{}_{ai}&=g_1{}_a{}^{a'} g_2{}_i{}^{i'} X_{a'i'},\  \bar X'{}_a^i=g_1{}_a{}^{a'}  \bar X^{i'}_{a'} g_2^{-1}{}_{i'}{}^{i},
\s]
where $g_1$ belongs to the defining representation of $O(N-1)$ in the subspace transverse to $\hat v$,
and $g_2$ belongs to the defining representation of $GL(R)$. 
Assuming that these symmetries do not break down in the large-$N$ limit, we can assume the following forms of
the two point functions:
\s[
 \langle \tbpsi{}_a^i \tpsi{}_{bj} \rangle &= Q_{11} I_{\perp\, ab}\, \delta^i_j, \\
 \langle \tbpsi{}_a^i \tvphi{}_{bj} \rangle &= Q_{12} I_{\perp\, ab}\, \delta^i_j ,   \\    
 \langle \tbvphi{}_a^i \tpsi{}_{bj} \rangle &= Q_{21} I_{\perp\, ab}\,\delta^i_j , \\  
 \langle \tbvphi{}_a^i \tvphi{}_{bj} \rangle &= Q_{22} I_{\perp\, ab}\, \delta^i_j, \\
 \hbox{Others}&=0,
 \label{eq:defofQ}
\s]   
where $Q_{ij}$ are some numbers.
From \eq{eq:defofseff} in \ref{app:sd}, the effective action written in terms of $Q_{ij}$ is given by
\s[
S_{\rm eff}&=
\langle K+V \rangle -(N-1)R \log(- \det Q )\\
&= (N-1)R \left(\epsilon Q_{11}-Q_{12}-Q_{21} -Q_{22} 
-\frac{|v|^2 (N-1)}{6 \alpha} \left(Q_{12} ^2 +Q_{21}^2 + 2 Q_{11} Q_{22}
\right)
-\log (- \det Q) \right)
\label{eq:seff}
\s]
 in the leading order of $N-1$.
Here the minus sign in the logarithm is appropriate for the current computation, 
since the solution of $Q_{ij}$ below gives $\det Q<0$. 
In the derivation of \eq{eq:seff}, we have used the factorizations into two-point functions and \eq{eq:defofQ}, such as 
\s[
\langle \tbpsi^i\cdot \tvphi{}_j \tbpsi^j\cdot \tvphi{}_i \rangle 
&=\langle \tbpsi^i{}_a \tvphi{}_{aj} \rangle{}  \langle \tbpsi^j{}_b \tvphi{}_{bi}\rangle-
\langle \tbpsi^i{}_a \tvphi{}_{bi}\rangle{}  \langle \tbpsi^j{}_b \tvphi{}_{aj} \rangle \\
&=(N-1)^2 R Q_{12}^2 - (N-1) R^2 Q_{12}^2,
\s]
and have taken only the leading order terms in $N-1$.

The stationary equations,
\[
\frac{\partial S_{\rm eff}}{\partial Q_{11}}=\frac{\partial S_{\rm eff}}{\partial Q_{12}}=\frac{\partial S_{\rm eff}}{\partial Q_{21}}
=\frac{\partial S_{\rm eff}}{\partial Q_{22}}=0,
\label{eq:stationary}
\]
have four independent solutions, and, among them, we should choose a solution which reproduces the $v=0$ limit 
(this is a free theory limit, since $V=0$) given by
\[
Q_{11}=\frac{1}{1+\epsilon},\ Q_{12}=Q_{21}=\frac{-1}{1+\epsilon},\ Q_{22}=\frac{\epsilon}{1+\epsilon}, \hbox{ for }v=0.
\label{eq:qveq0}
\]
The explicit expression of the solution is given in \ref{app:stationary} with $x=v^2 (N-1)/(3 \alpha)$. 
The $\epsilon\rightarrow +0$ limit has a singularity at $x=1/4$, and for each region, we obtain for $\epsilon\rightarrow +0$
\begin{itemize}
\item $0<x<1/4$
\s[
&Q_{11}=\frac{-\sqrt{1-4 x}+1}{2 x\sqrt{1-4 x}},\\
&Q_{12}=Q_{21}=\frac{1-\sqrt{1-4 x}-4 x}{2 x \sqrt{1-4 x}},\\
& Q_{22}=0, 
\label{eq:qvalsmall}
\s]
\item $x>1/4$
\s[
&Q_{11}=\frac{\sqrt{-1+4 x}}{2 \sqrt{\epsilon} x}-\frac{1}{2 x}+{\cal O}(\sqrt{\epsilon}), \\
&Q_{12}=Q_{21}=-\frac{1}{2 x}+\frac{\sqrt{\epsilon}}{2 x \sqrt{-1+4 x}}+{\cal O}(\epsilon^\frac{3}{2}),\\
&Q_{22}=-\frac{\sqrt{\epsilon} \sqrt{-1+4 x}}{2 x}+\frac{\epsilon}{2 x} +{\cal O}(\epsilon^\frac{3}{2}).
\label{eq:qvallarge}
\s]
\end{itemize}
The latter solution is not well-defined in $\epsilon\rightarrow +0$, but the $S_{\rm eff}$ for those solutions has well-defined limits
in both regions:
\[
S^{\epsilon=+0}_{\rm eff}(x)=\left\{
\begin{array}{ll}
\frac{(N-1)R}{2} \left(2+\log(16)+\frac{1-\sqrt{1-4 x}}{x}-4 \log\left(1-\sqrt{1-4 x}\right)+4 \log(x) \right)& \hbox{ for } 0<x<\frac{1}{4}, \\
\frac{(N-1)R}{2 x}(1+2 x+2 x \log(x))& \hbox{ for } \frac{1}{4}<x.
\end{array}
\right.
\label{eq:seffzero}
\]
Then from \eq{eq:appzeff} we obtain for large-$N$
\[
\rho(v,R,+0)=3^\frac{N-1}{2} \pi^{-\frac{N}{2}} \alpha^\frac{N}{2} |v|^{-2 N} e^{-\frac{\alpha}{|v|^2}} 
e^{S^{\epsilon=+0}_{\rm eff}(x)-2 (N-1) R},
\label{eq:rhosemi}
\]
where the subtraction in the exponent is due to $S^{\epsilon=+0}_{\rm eff}(0)=2 (N-1) R$.
Note that, in the above computation,  
the limiting process of $\epsilon\rightarrow +0$ is independent from taking $R=1/2$, 
and therefore the result does not depend on the order of them.

A delicate matter concerning taking $R=1/2$ is that $R$ in the above computation
is assumed to be positive integers, since $R$ is the number of replicas. 
Therefore we do not know whether this expression can truly be analytically extended to $R=1/2$.
This is generally a difficult question to answer, which is often raised in applying the replica trick. 
 However, as is shown in  \ref{app:relation}, the above result \eq{eq:rhosemi} with \eq{eq:seffzero} 
 can be shown to agree with
 a large-$N$ expression computed by random matrix theory in another context.
 
Finally, we want to argue that only the expression at $x>1/4$ in \eq{eq:seffzero} practically matters, 
when $N$ is large. The reason is as follows. 
Because of the exponential damping factor $e^{-\frac{\alpha}{|v|^2}}$,  $\rho(v,R,+0)$ takes significant values only in the region
$|v|\gtrsim \sqrt{\alpha}$. This means that the relevant region satisfies
$x=|v|^2 (N-1)/(3 \alpha) \gtrsim (N-1)/3>1/4$ for large $N$. 
Therefore, we can practically use $S_{\rm eff}^{\epsilon=+0}$ at $x>1/4$ for all the region of $x$ for large-$N$.
By doing this, we finally obtain, for large-$N$ and $R=1/2$,
\[
\rho(v)=\rho(v,1/2,+0)=(N-1)^{\frac{N-1}{2}} e^{-\frac{N-1}{2}} \pi^{-\frac{N}{2}} \alpha^{\frac{1}{2}}|v|^{-N-1} e^{-\frac{\alpha}{4 |v|^2} }.
\label{eq:rhofinal}
\]

The result \eq{eq:rhofinal} is for the distribution of vector $v$. For comparison with Monte Carlo simulations, it is more convenient 
to transform it to the distribution of the size $|v|$. Since $d^Nv=S_{N-1} |v|^{N-1}d|v|$ with $S_{N-1}=2 \pi^{N/2}/\Gamma(N/2)$ being 
the surface volume of a unit sphere in $N$-dimensions, we obtain for large-$N$
\[
\rho_{\rm size}(|v|)=\frac{2 (N-1)^{\frac{N-1}{2}} e^{-\frac{N-1}{2}} \alpha^{\frac{1}{2}}}{\Gamma\left(\frac{N}{2}\right)}
|v|^{-2} e^{-\frac{\alpha}{4 |v|^2} }.
\label{eq:rhosize}
\] 

One can also transform the result to the real eigenvalue distribution (Z-eigenvalues in the terminology of \cite{Qi}), using \eq{eq:rhoeigen}: 
For large-$N$,
\[
\rho_{\rm eigevalue}(z)=\frac{2 (N-1)^{\frac{N-1}{2}} e^{-\frac{N-1}{2}} \alpha^{\frac{1}{2}}}{\Gamma\left(\frac{N}{2}\right)}
e^{-\frac{\alpha}{4} z^2}. 
\label{eq:rhoeig}
\] 
Interestingly, the real eigenvalue distribution is just given by the Gaussian function in this approximation.

\section{Comparison with Monte Carlo simulations}
\label{sec:monte}
In this section, we perform some Monte Carlo simulations, and compare the results with the exact small-$N$ and 
the approximate large-$N$ results, which are respectively obtained in Sections~\ref{sec:exact} and \ref{sec:approximation}.

The eigenvector equations \eq{eq:eg} are a system of polynomial equations, which can be solved by appropriate 
polynomial equation solvers. We use Mathematica 12 for this purpose.
It generally produces all the solutions including complex ones, and we pick up all the real ones out of them.  
Whether all the real eigenvectors are covered can be guaranteed by 
checking whether the number of all the generally complex solutions agrees with the known formula, 
$2^N-1$ \cite{cart}. 

With the above way of obtaining real eigenvectors, our procedure of numerical simulations can be summarized as follows.
\begin{itemize}
\item
Randomly generate $C$ with the Gaussian distribution: Each independent component is randomly 
generated by $C_{ijk}=\sigma_{ijk}/\sqrt{d(i,j,k)}, \ 
(i\leq j \leq k)$. Here $\sigma_{ijk}$ is a random number following the normal distribution of mean value zero and standard 
deviation one. $d(i,j,k)$ is a degeneracy factor defined by 
\s[
d(i,j,k)=\left\{ 
\begin{array}{ll}
1, & i=j=k \\
3 ,& i=j\neq k ,\,i\neq j=k,\, k=i\neq j \\
6 ,& i\neq k \neq j \neq i
\end{array}
\right.
.
\s]
The above distribution corresponds to choosing $\alpha=1/2$ in \eq{eq:rhov}, since
\s[
C^2=C_{abc}C_{abc}=\sum_{i,j,k=1 \atop i\leq j \leq k}^N d(i,j,k) \, C_{ijk} ^2
\s]
due to $C$ being a symmetric tensor.
\item
Pick up all the real eigenvectors by the way mentioned above.
\item
Store the size $|v|$ and the value of $\det M$ for each of all the real eigenvectors.
\item 
Repeat the above processes.
\end{itemize}
By the above repeating procedure we obtain a data of $(|v_i|,\det M_i) \ (i=1,2,\ldots,L)$, where $L$ is the total number of 
real eigenvectors generated. For the general case of $R$, we define
\[
\rho_{R}^{\rm sim} ((k+1/2)\delta v)=\frac{1}{ \delta v N_C}\sum_{i=1}^L  |\det M|^{2R-1} \,
\theta(k \delta v <|v_i|\leq (k+1)\delta v),
\label{eq:rhor}
\]
where $N_C$ denotes the total number of randomly generated $C$, $\delta v$ is a bin size, $k=0,1,2,\ldots$, and $\theta$ is a support function which takes 1 if the inequality of the argument
is satisfied, but zero otherwise. 
This quantity corresponds to $\rho(v,R,0) S_{N-1} |v|^{N-1}$ with $\alpha=1/2$ of the analytical result
by a derivation similar to that of \eq{eq:rhosize}.
The $-1$ in the exponent of $|\det M|$ in \eq{eq:rhor} comes from the consideration of the measure 
associated to the delta functions, namely, the difference between \eq{eq:defrhoC} and \eq{eq:defrho}.

\begin{figure}
\begin{center}
\hfil
\includegraphics[width=4cm]{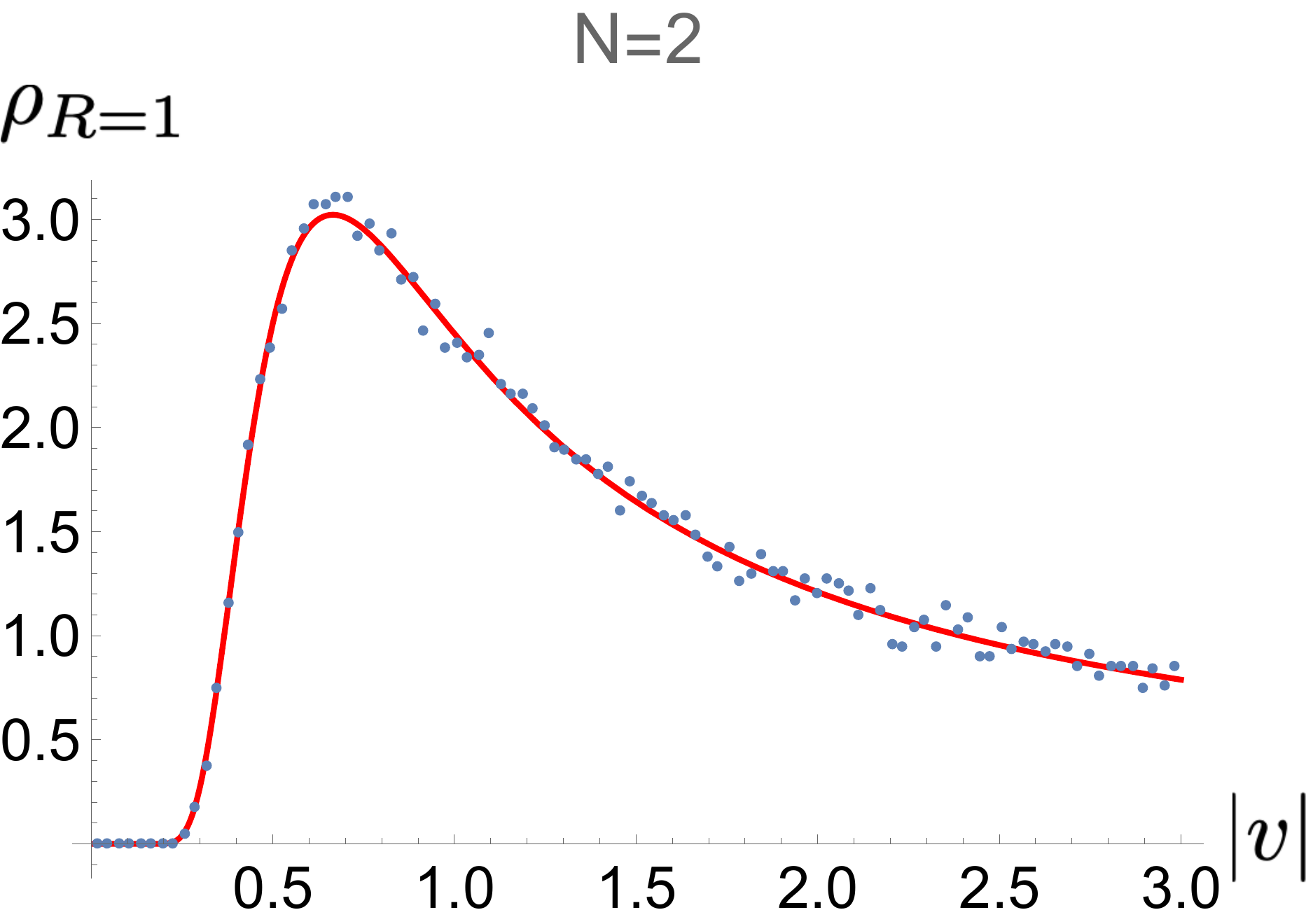}
\hfil
\includegraphics[width=4cm]{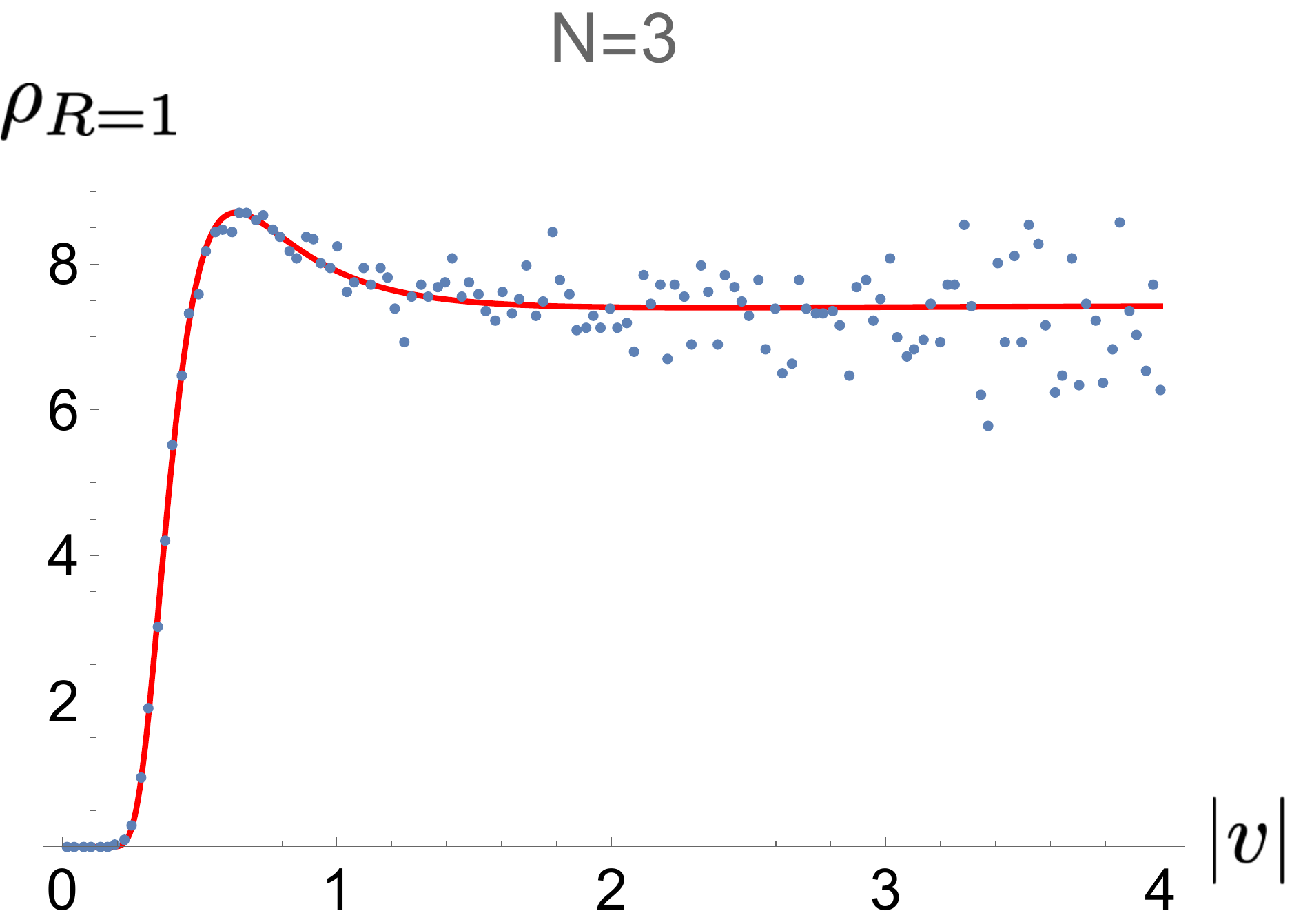}
\hfil
\includegraphics[width=4cm]{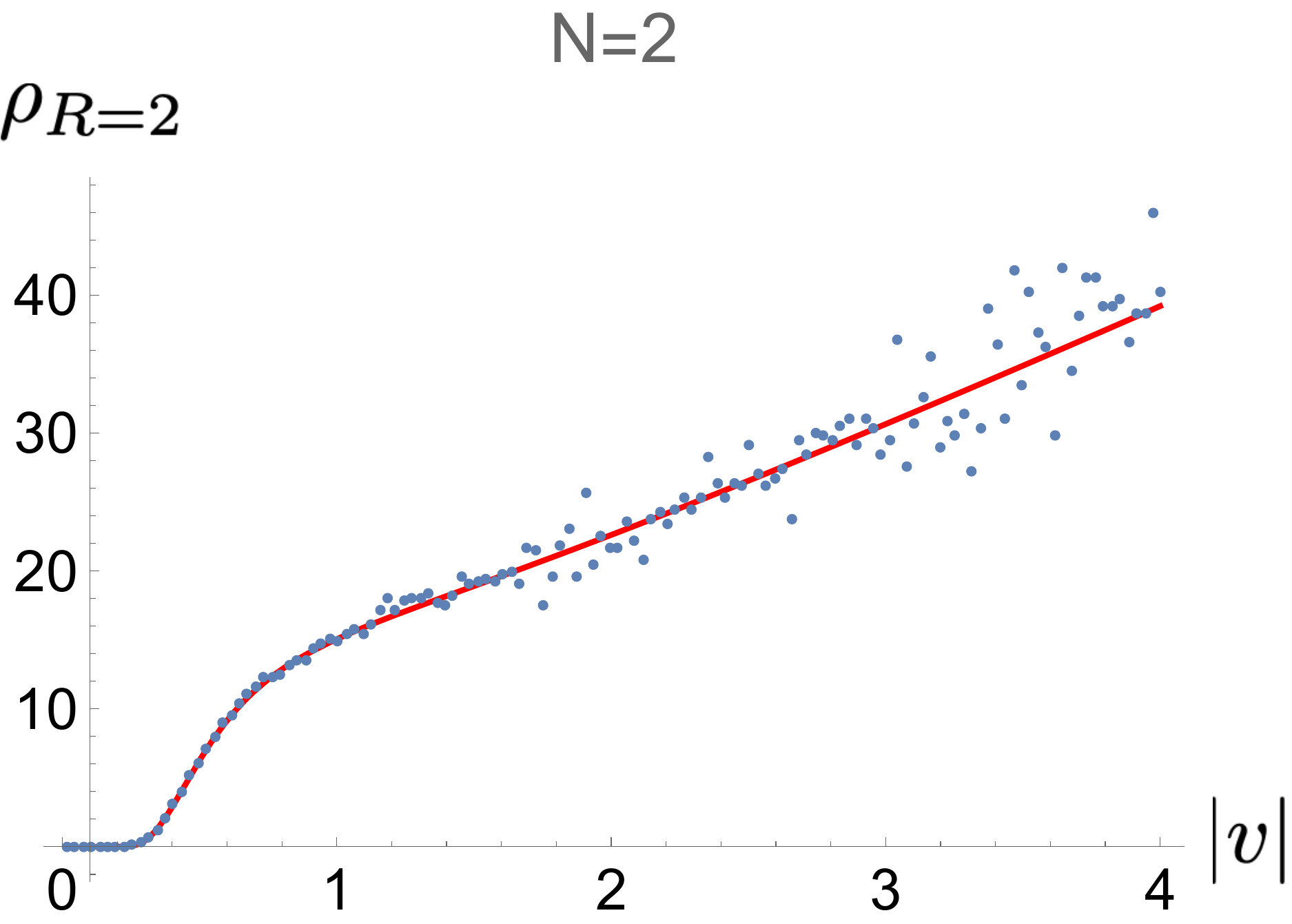}
\hfil
\includegraphics[width=4cm]{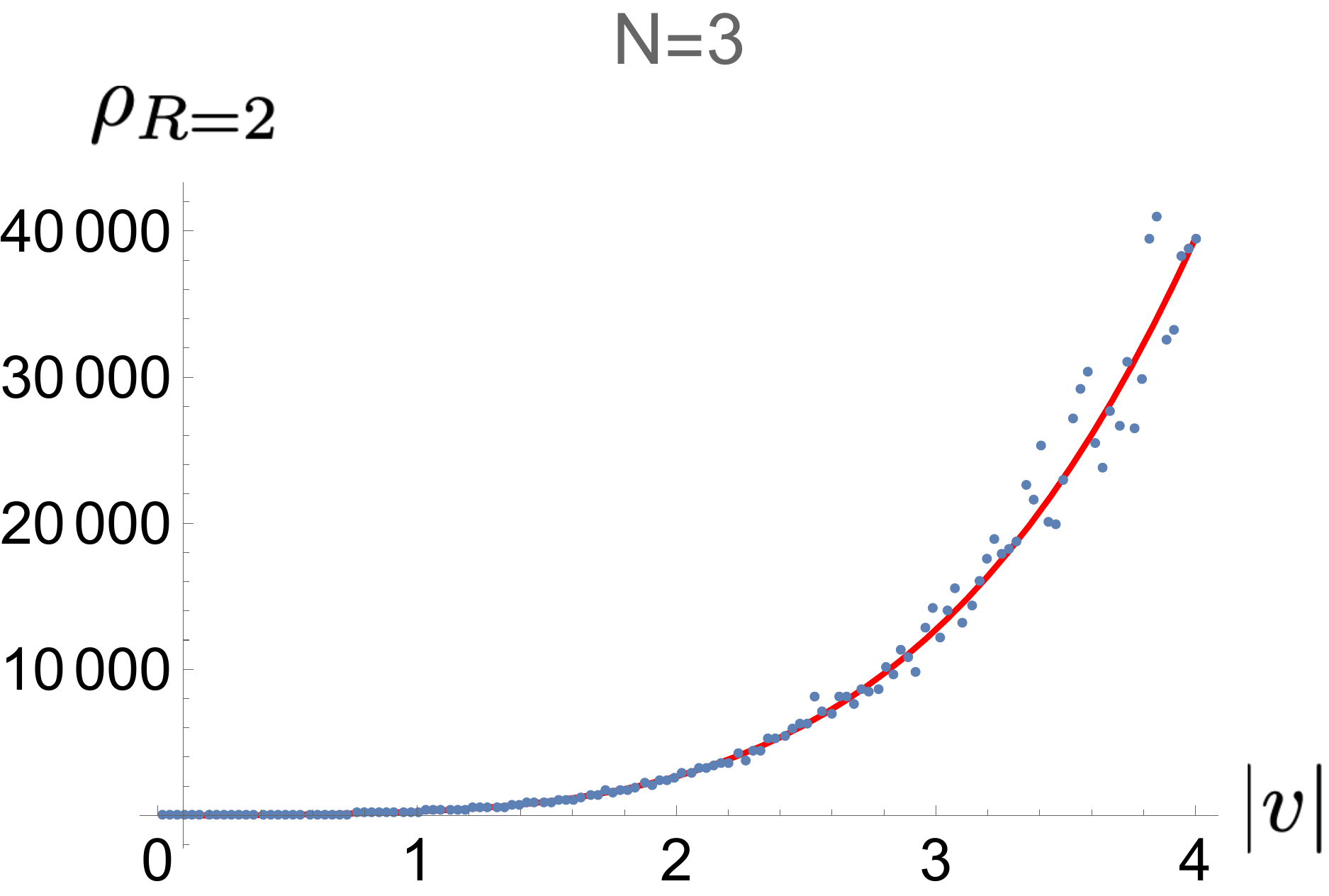}
\hfil
\caption{
Some non-trivial checks of the general formula \eq{eq:fourfermi} with  \eq{eq:vint}.
The analytical results,  \eq{eq:rhol} 
with \eq{eq:nreq} for $\alpha=1/2$, (solid lines) and the results from the Monte Carlo simulations  
\eq{eq:rhor} (dots) for $(N,R)=(2,1)$, 
$(3,1)$, $(2,2)$, $(3,2)$ from the left to the right panels are compared. 
$\delta v=0.03$. $N_C=3\cdot 10^4$ for the former two, and 
$N_C=10^5$ for the latter two.
\label{fig:neq2}
}
\end{center}
\end{figure} 

In Figure~\ref{fig:neq2}, the numerical datas \eq{eq:rhor} for $N=2,3, \ R=2,3$, and the analytical 
results, $\rho(v,R,0) S_{N-1} |v|^{N-1}$ with $\rho(v,R,0)$ being given by \eq{eq:rhol} 
with \eq{eq:nreq} for $\alpha=1/2$, are compared. 
They precisely agree. The agreement includes the allover numerical factor, and gives non-trivial checks
of the general formula \eq{eq:fourfermi} with \eq{eq:vint}.

For $R=1/2$ in \eq{eq:rhor}, we have
\[
\rho_{\rm size}^{\rm sim} ((k+1/2)\delta v)=\frac{1}{ \delta v N_C}\sum_{i=1}^L \theta(k \delta v <|v_i|\leq (k+1)\delta v).
\label{eq:rhosim}
\]
For large-$N$, this is the numerical quantity corresponding to \eq{eq:rhosize} with $\alpha=1/2$.

\begin{figure} 
\begin{center}
\includegraphics[width=7cm]{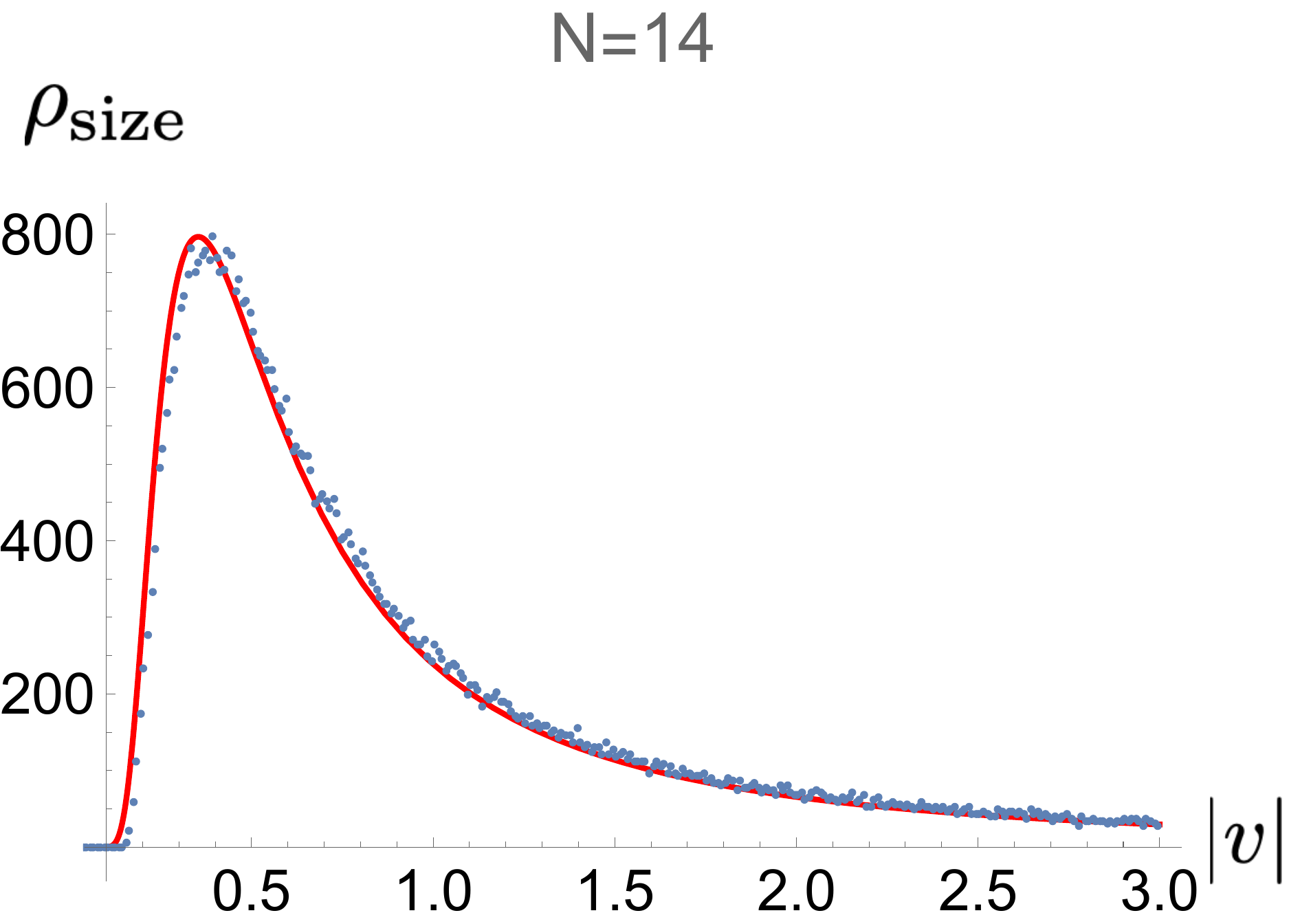}
\hfil
\includegraphics[width=7cm]{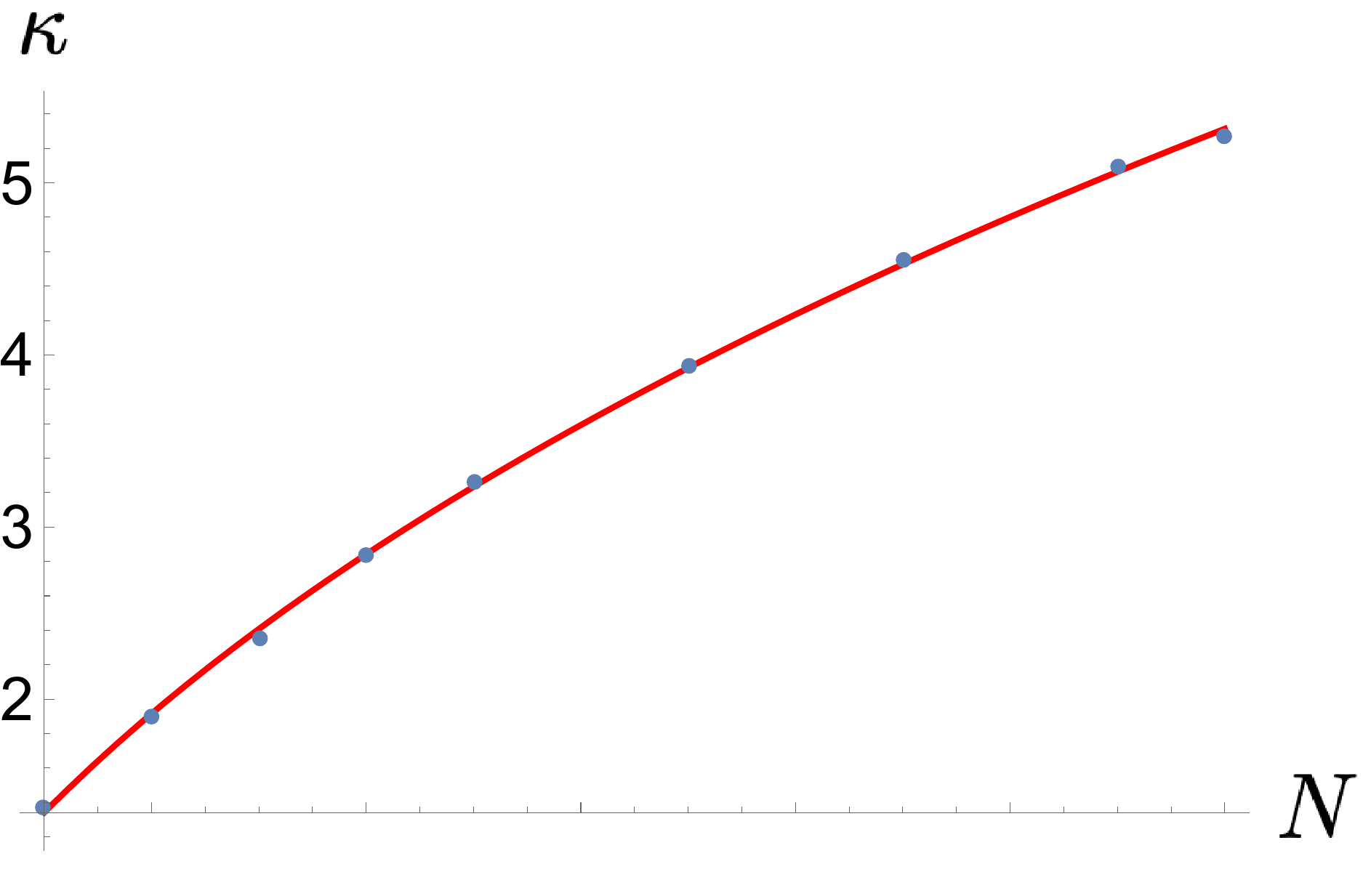} 
\caption{\label{fig:v} 
Left: \eq{eq:rhosim} from a Monte Carlo simulation for $N=14$ is plotted by dots. $N_C=400,\ \delta v=0.01$. 
This is compared with $\kappa \rho_{\rm size}$ from \eq{eq:rhosize} with $\kappa=5.27$ (the best fitting value), which is plotted by a solid line.
Right: The overall factors $\kappa$ needed for the best fitting for various $N$ are plotted by dots. 
The fitting line is $-3.6+3.2 N^{0.38}$.}
\end{center}
\end{figure}

In the left panel of Figure~\ref{fig:v} the analytical result \eq{eq:rhosize} 
and the numerical result \eq{eq:rhosim} are compared. 
In this figure, the analytical result is multiplied by an extra allover numerical factor, 
which means that the overall factor is not 
correctly computed in the analytical result, while the functional form agrees well with the numerical data. 
As shown in the right panel, the extra factor $\kappa$ needed to have good agreement is dependent on $N$. 
As far as the fitting line implies, the factor seems to asymptotically diverge in $N\rightarrow \infty$.

\begin{figure}
\begin{center}
\includegraphics[width=7cm]{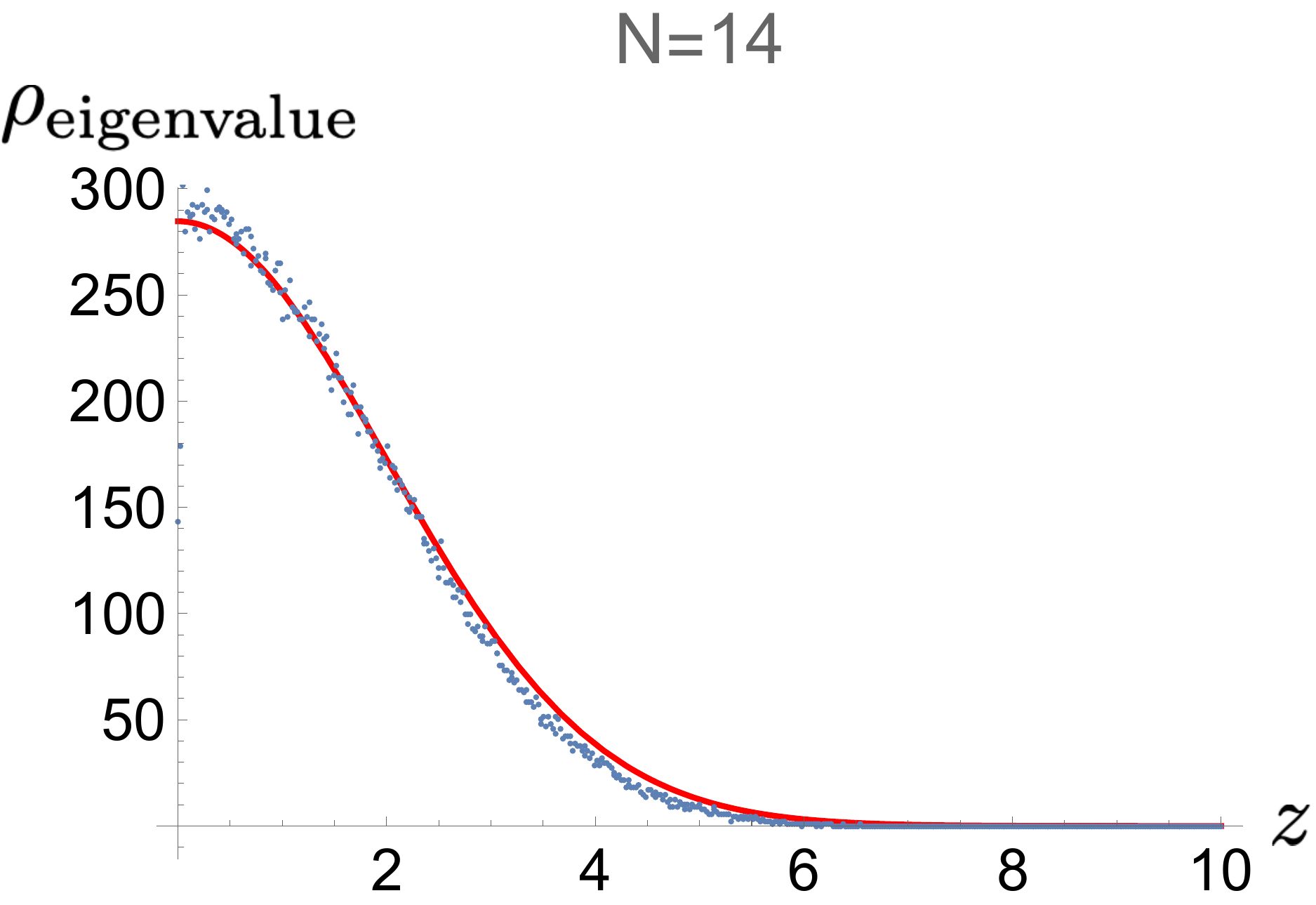}
\caption{\label{fig:z} 
The numerical result \eq{eq:rhosimz} and the analytical result \eq{eq:rhoeig} for large-$N$ 
are compared for $N=14$.
$N_C=400,\ \delta v=0.02$, and $\kappa=5.6$ (the best fitting value).}
\end{center}
\end{figure}

In a similar manner, one can obtain the real eigenvalue (Z-eigenvalue) distribution from the numerical data
by using the relation $z=1/|v|$ mentioned in a sentence above \eq{eq:rhoeigen}:
\[
\rho_{\rm eigenvalue}^{\rm sim} ((k+1/2)\delta z)=\frac{1}{\delta z N_C}\sum_{i=1}^L \theta\left(
k \delta z <\frac{1}{|v_i|}\leq (k+1)\delta z\right).
\label{eq:rhosimz}
\]
For large-$N$ this corresponds to the analytic expression \eq{eq:rhoeig}, and 
they are compared in Figure~\ref{fig:z}. 
We again find good agreement on the functional form, which is simply the Gaussian.

The mismatch of the overall factor may be corrected 
by the next-leading order computation. As shown in \eq{eq:qvallarge},
$Q_{ij}$ are in the order of $1/\sqrt{N}$. Then the next-leading order corrections would be $\delta Q_{ij} \sim 1/N$ or $1/N^{3/2}$. 
Putting this correction to \eq{eq:seff} can produce a substantial correction to $S^{\epsilon=+0}_{\rm eff}$, which can change the 
overall factor. There are also some other potential sources of corrections,  
and it will be a complex task to compute all of them, because of the many interaction terms of $V$ in \eq{eq:vint}.

\section{Summary and future problems}
Considering the important roles of the eigenvalue distributions in matrix models, 
it would be interesting to compute corresponding
quantities in tensor models and study their properties and roles.  
In this paper, we have studied the real tensor eigenvalue/vector distributions for random real symmetric order-three tensors
with the Gaussian distribution. 
We first rewrote the problem as the computation of a partition function of
a four-fermi theory with $R$ replicated fermions. 
We exactly computed the partition function for some small-$N,R$ cases, and 
found precise agreement with the Monte Carlo results.
For large-$N$ we approximately computed the partition function 
by using a Schwinger-Dyson equation for the two-point functions. 
Interestingly the real eigenvalue distribution obtained by taking $R=1/2$ 
turned out to be simply the Gaussian within this approximation.
The comparison with the Monte Carlo simulations showed that, by multiplying  an overall numerical factor depending on $N$,
the approximate analytic expression was in good agreement with the numerical results.

An obvious future problem is to improve the approximate computation of the partition function of the four-fermi theory 
for large-$N$ so that the overall factor is also correctly derived.
In this paper, we used a Schwinger-Dyson equation for the two-point functions, but it is feasible to extend it to 
include higher-point functions, though this will be much more complicated because of the presence of many terms of the four-fermi 
interactions. Another possible direction is that, 
while we considered a replicated fermion system to take analytic continuation to the wanted case, 
it is also possible to consider a fermion-boson system to directly compute the distribution without analytic continuation. 
It will be useful for future extended study to figure out which ways of computations give correct results efficiently.
It also seems important to check whether the simple Gaussian distribution of the real eigenvalues is robust against the improvement.

Another interesting direction will be to extend the Gaussian tensor model we considered to more general tensor models.
In particular, an interesting phenomenon which occurs in matrix models is 
the topological change of eigenvalue distributions. 
As for tensor models, first of all, it is unknown whether there exists such a topological transition
in tensor eigenvalue/vector distributions, and, if it exists, how this is 
related to the dynamics of tensor models is also an interesting question. 
It would be worth mentioning that a similar phenomenon with topological changes of distributions 
of dynamical variables was observed in the analysis of the wave function of a tensor model in the 
Hamiltonian formalism \cite{Kawano:2021vqc,Sasakura:2022aru}.
We would therefore expect that such phenomena could be universal across matrix and tensor models, and 
this could be studied by extending the current work to more general cases.  
   
\vspace{.3cm}
\section*{Acknowledgements}
The author is supported in part by JSPS KAKENHI Grant No.19K03825. 

\appendix
\def\thesection{Appendix \Alph{section}}

\section{Approximate computations with the Schwinger-Dyson equation}
\label{app:sd}
In this appendix, we explain the approximation using the Schwinger-Dyson equation, 
which we employ in the text. 
The content of this section is rather standard, and is therefore supposed to be for some unfamiliar readers.

The problem we are dealing with is a four-fermi theory which can generally be written in the form,
\[
S=\bar \psi_a K_{ab} \psi_b + g B_{abcd} \bar \psi_a\bar \psi_b \psi_c \psi_d,
\]
where $g$ is a coupling, and $B$ is assumed to satisfy the partially antisymmetry condition given by
\[
B_{abcd}=-B_{bacd}=-B_{abdc},
\]
due to the fermionic property of the variables.
The action $S$ is invariant under a scaling transformation,
\[
\bar \psi_a'= s \bar \psi_a, \ \psi_a'= s^{-1} \psi_a,
\label{eq:scale}
\]
where $s$ can be an arbitrary number.
The partition function is defined by
\[
Z=\int d\bar\psi d\psi\, e^S,
\]
where the integration measure is defined so that \cite{zinn}
\[
\int d\bar\psi d\psi\, e^{\bar \psi_a K_{ab} \psi_b}=\det K.
\]

Since an integral of a total derivative identically vanishes for the fermionic integration, 
we can write down the following consistency equation,
\s[
0&=\int d\bar \psi d\psi \frac{\partial}{\partial \bar \psi_a}( \bar \psi_b \, e^S) \\
&=\int d\bar \psi d\psi \left( \delta_{ab} -K_{ac} \bar \psi_b \psi_c -2  g B_{acde} \bar \psi_b \bar \psi_c \psi_d \psi_e\right)e^S,
\s]
where the minus signs are due to the fermionic property.
By dividing by $Z$, we obtain a Schwinger-Dyson equation, 
\s[
K_{ac} \langle \bar \psi_b \phi_c \rangle +2 g B_{acde} \langle \bar \psi_b  \bar \psi_c \psi_d \psi_e \rangle=\delta_{ab},
\label{eq:appsd}
\s]
where $\langle {\cal O} \rangle$ denotes correlation functions defined by
\[
\langle {\cal O}\rangle =\frac{1}{Z} \int d\bar\psi d\psi\, {\cal O} e^S.
\]

The four-point function in \eq{eq:appsd} may be expanded in terms of connected correlation functions,
\[
\langle \bar \psi_b  \bar \psi_c \psi_d \psi_e \rangle=-\langle \bar \psi_b   \psi_d  \rangle \langle \bar \psi_c \psi_e \rangle
+\langle \bar \psi_b  \psi_e  \rangle \langle \bar \psi_c  \psi_d  \rangle+\langle \bar \psi_b  \bar \psi_c \psi_d \psi_e \rangle_c,
\label{eq:fourpoint}
\]
where $\langle {\cal O} \rangle_c$ denotes the connected correlation functions.
Here we have assumed that only the correlation functions with the same number of $\bar \psi$ and $\psi$ remain non-zero
due to the symmetry \eq{eq:scale}. Now assuming that we can ignore the four-point connected
correlation functions as higher-order terms, $\langle \bar \psi_b  \bar \psi_c \psi_d \psi_e \rangle_c\simeq 0$,
\eq{eq:appsd} and \eq{eq:fourpoint} lead to a consistency equation for the two-point functions,
\[
K_{ac} \langle \bar \psi_b \phi_c \rangle +4 g B_{acde} 
\langle \bar \psi_b  \psi_e  \rangle \langle \bar \psi_c  \psi_d  \rangle=\delta_{ab}.
\label{eq:consist1}
\]
In fact, the equation \eq{eq:consist1} can be written in another way. Let us denote $G_{ab}=\langle \bar \psi_a \psi_b \rangle$.
Then, by applying $G^{-1}$ on both sides of \eq{eq:consist1}, we obtain
\[
K_{ab} +4 g B_{acdb} G_{cd}=G^{-1}_{ba}.
\label{eq:consist2} 
\]
By solving this equation one can determine $G$. 

Once we know $G$, the partition function $Z$ can be computed in the following manner.
We first note that
\s[
\frac{d}{d g} \log Z&= B_{abcd} \langle \bar \psi_a  \bar \psi_b \psi_c \psi_d \rangle \\
&=2  B_{abcd} G_{ad} G_{bc} 
\label{eq:dzdg}
\s]
by applying the approximation taken above. 
Therefore one can determine $Z$ by solving the equation \eq{eq:dzdg} 
with the initial condition,
\[
Z(g=0)=\det K.
\]

The procedure above can be summarized as a more intuitive process.
Let us define
\s[
S_{\rm eff}&=\langle S \rangle -\log \det G \\
&=K_{ab} G_{ab}+2 g B_{abcd} G_{ad} G_{bc} - \log \det G,
\label{eq:defofseff}
\s]
where $\langle S \rangle$ in \eq{eq:defofseff} has been computed with the approximation taken above. 
Here it may be more appropriate to replace $\log \det G$ with $\log(-\det G)$, depending on the sign of $\det G$ of a solution. 
In \eq{eq:seff} it is chosen to be the latter because the solution turns out to give $\det G<0$. 
Anyway, the difference is a constant and hence does not essentially affect the discussions below.

Then we consider the stationary condition with respect to $G$:
\s[
0&=\frac{\partial}{\partial G_{ab} }S_{\rm eff} \\
&= K_{ab} +4 g B_{acdb} G_{cd} -G^{-1}_{ba}.
\label{eq:appstationary}
\s]
This indeed agrees with \eq{eq:consist2}.

Moreover, let us define 
\[
Z^0_{\rm eff}=\det K\,  e^{S^0_{\rm eff}-S_{\rm eff}^0(g=0)},
\label{eq:appzeff}
\]
where $S^0_{\rm eff}$ is defined by inserting the solution $G=G^0$ of \eq{eq:appstationary} (or \eq{eq:consist2})
into  \eq{eq:defofseff}. 
Then we can find that
\[
\frac{d}{dg}\log Z^0_{\rm eff}=\frac{\partial S^0_{\rm eff}}{\partial g}  + \frac{\partial S^0_{\rm eff}}{\partial G^0_{ab} }\frac{d G^0_{ab}}{dg}=
2 g B_{abcd} G^0_{ad} G^0_{bc},
\label{eq:derz}
\]
where we have used the stationary condition \eq{eq:appstationary}.
Indeed \eq{eq:derz} agrees with \eq{eq:dzdg}. We also find $Z^0_{\rm eff}(g=0)=\det K$. 
Therefore, $Z^0_{\rm eff}$ gives the partition function
in this approximation.

\section{The solution to \eq{eq:stationary}}
\label{app:stationary}
The solution to \eq{eq:stationary} reproducing \eq{eq:qveq0} at $v=0$ is given by
\s[
&Q_{11}=\frac{-2+\sqrt{2} \sqrt{\frac{-1+\epsilon+4 x+\sqrt{(1+\epsilon-4 x)^2+16 \epsilon x}}{\epsilon}}}{4 x},\\
&Q_{12}=Q_{21}\\
&\ \ \ \ \ =\frac{1}{8x}\left(-4+\sqrt{2} \sqrt{\frac{-1+\epsilon+4 x+\sqrt{(1+\epsilon-4 x)^2+16 \epsilon x}}{\epsilon}} \right.\\
&\hspace{2cm}-\sqrt{2} \epsilon \sqrt{\frac{-1+\epsilon+4x+\sqrt{(1+\epsilon-4 x)^2+16 \epsilon x}}{\epsilon}} \\
&\hspace{2cm}-4\sqrt{2} x \sqrt{\frac{-1+\epsilon+4 x+\sqrt{(1+\epsilon-4 x)^2+16 \epsilon x}}{\epsilon}} \\
&\hspace{2cm}\left.+\sqrt{2} \sqrt{(1+\epsilon-4 x)^2+16 \epsilon x} \sqrt{\frac{-1+\epsilon+4 x+\sqrt{(1+\epsilon-4 x)^2+16 \epsilon x}}{\epsilon}}\right), \\
&Q_{22}=-\frac{\epsilon \left(-2+\sqrt{2} \sqrt{\frac{-1+\epsilon+4 x+\sqrt{(1+\epsilon-4 x)^2+16 \epsilon x}}{\epsilon}}\right)}{4 x}.
\s]

\section{Relation to the spherical $p$-spin model}
\label{app:relation}
In this appendix, we explain the relation between the tensor eigenvalue problem and the computation of 
the complexity\footnote{See for example \cite{example} for a review.} 
in the spherical $p$-spin model \cite{pspin,pedestrians}. 
In the end, we will find that \eq{eq:seffzero} agrees with 
the large-$N$ result given in \cite{randommat}, which was computed by random matrix theory.
In this appendix, we limit ourselves to $p=3$, following our limited setting.

The Hamiltonian of the spherical $p$-spin model (with $p=3$) is given by 
\[
H=-\frac{1}{N} J_{abc} \sigma_a \sigma_b \sigma_c, 
\label{eq:hamiltonian}
\]
with the continuous spin variable $\sigma \in \mathbb{R}^N$ which is constrained by 
\[
\sigma_a \sigma_a =N.
\label{eq:normalsigma}
\]
Here the symmetric real tensor $J$ has the standard Gaussian distribution and can be identified with
$C$ with $\alpha=1/2$ in our notation. To match \eq{eq:normalsigma} with our notation, we perform
\[
\sigma_a=\sqrt{N} w_a,
\]
and then we have
\[
H=-\sqrt{N} C_{abc} w_a w_b w_c \hbox{ with } w_a w_a=1.
\label{eq:hampspin}
\]

The problem of the complexity of the $p$-spin spherical model is 
to count the number of local minima (and stationary points) of the 
above Hamiltonian.
By using the method of Lagrange multiplier, counting the stationary points is equivalent to solve 
\[
\sqrt{N} C_{abc} w_b w_c=-N u w_a,
\label{eq:egham}
\]
where $u\in \mathbb{R}$, and the factor of $N$ and the minus sign on the righthand side are 
for later convenience. 
Physically $u$ can be interpreted as the mean energy per site, since $u=H/N$ at the 
stationary points. \eq{eq:egham} is indeed the same as the Z-eigenvalue equation 
\eq{eq:zeigeq} by the identification,
\[
z=-\sqrt{N} u.
\label{eq:zu}
\]

Now one of the main results stated in \cite{randommat} is 
\[
\lim_{N\rightarrow \infty} \frac{1}{N} \log \mathbb{E} \hbox{Crt}_N(u) =\Theta_{p}(u),
\label{eq:limtheta}
\] 
where 
\[
\Theta_p(u)=\left\{
\begin{array}{ll}
\frac{1}{2}\log(p-1) -\frac{p-2}{4 (p-1)} u^2 -I_1 (u), & \hbox{if }u<-E_{\infty}, \\
\frac{1}{2} \log(p-1) -\frac{p-2}{4 (p-1)} u^2, &  \hbox{if }-E_{\infty}\leq u < 0, \\
\frac{1}{2} \log(p-1),& \hbox{if } 0\leq u. 
\end{array}
\right.
\]
Here $\mathbb{E} \hbox{Crt}_N(u)$ denotes the expectation value of the number of the critical (stationary) 
points below energy $N u$ of the Hamiltonian \eq{eq:hamiltonian}, $E_\infty=2 \sqrt{(p-1)/p}$, and 
\[
I_1(u)=-\frac{u}{E^2_\infty} \sqrt{u^2-E^2_\infty}-\log \left( -u+\sqrt{u^2-E^2_\infty}\right) + \log E_\infty.
\]
Since $\mathbb{E} \hbox{Crt}_N(u) \sim e^{N \Theta_{p}(u)}$ from \eq{eq:limtheta}, the distribution of critical
points would be given by $\rho(u)du \sim  d (e^{N \Theta_{p}(u)})\sim e^{N \Theta_{p}(u) } du$
in the leading large-$N$ order of the exponent.

On the other hand, the corresponding distribution in our computation is given by 
\[
\rho(v,R=1/2,+0) \,S_{N-1} v^{N-1}  dv,
\label{eq:comparev}
\]
where $v=|v|$, the volume factor has been multiplied, $S_{N-1}=2 \pi^{N/2}/\Gamma(N/2)$, and 
$\rho(v,R,+0)$ is given in \eq{eq:rhosemi} with \eq{eq:seffzero}. 
From \eq{eq:zu} and $z=1/v$ (see a sentence below \eq{eq:zeigeq}),
our parameters are related to $u$ by
\[
v=-\frac{1}{\sqrt{N} u },
\]
and hence $x=2/(3 u^2)$.
Putting them into \eq{eq:comparev}, it indeed agrees with $e^{N \Theta_{p=3}(u) } du$ in the leading large-$N$
order of the exponent.

\end{document}